\documentclass[runningheads]{llncs}

\usepackage{graphicx}        
\usepackage{hyperref}
\usepackage{graphicx}
\usepackage{color}
\usepackage{amssymb}
\usepackage{pgfplots}
\pgfplotsset{compat=1.13}
\usepackage[normalem]{ulem}
\usepackage{tikz}

\newcommand{\OnlyLongVersion} [1] {#1}
\newcommand{\OnlyShortVersion} [1] {}

\begin{document}
\title{Balanced Schnyder woods for planar triangulations: an experimental study with applications
to graph drawing and graph separators\thanks{This work is supported by the French ANR GATO (ANR-16-CE40-0009-01).}}

\titlerunning{Balanced Schnyder woods for planar triangulations}

\author{Luca Castelli Aleardi~\inst{1}\orcidID{0000-0002-1142-2562}
}

\authorrunning{L. Castelli Aleardi}

\institute{LIX, Ecole Polytechnique, Institut Polytechnique de Paris, France \email{amturing@lix.polytechnique.fr}}

\maketitle              

\begin{abstract}
In this work we consider balanced Schnyder woods for planar graphs, which
are Schnyder woods where the number of incoming edges of each color
at each vertex is balanced as much as possible.
We provide a simple linear-time heuristic leading to obtain well balanced Schnyder woods in practice.
As test applications we consider two important algorithmic problems: the computation of Schnyder drawings 
and of small cycle separators.
While not being able to provide theoretical guarantees, our experimental results (on a wide collection of planar graphs) 
suggest that the use of balanced
Schnyder woods leads to an improvement of the quality of the layout of Schnyder drawings,
and provides an efficient tool for computing short and balanced cycle separators.
\end{abstract}


\section{Introduction}\label{sec:intro}

Schnyder woods~\cite{Schnyder90} and its generalizations are a deep tool for dealing with 
the combinatorics of planar~\cite{Felsner04} and surface maps~\cite{AleardiFL09,GoncalvesL14,DespreGL17}.
They lead to efficient algorithmic and combinatorial solutions for a broad collection of problems,
arising in several domains, from enumerative combinatorics to graph drawing and computational geometry.
For instance, the use of Schnyder woods has led to linear-time algorithms for grid drawing~\cite{Schnyder90,BonichonFM07,GoncalvesL14}, to the optimal
encoding and uniform sampling of planar maps~\cite{Pou03}, to the design of compact data structures~\cite{AleardiD18} and
to deal with geometric spanners~\cite{BonichonGHP10}.
Schnyder woods lead to fast implementations (also integrated in open source libraries\cite{Pigale}) 
and provide strong tools for establishing rigorous theoretical
guarantees that hold in the worst case, even for irregular, random or pathological cases.
The main idea motivating this work is that, in practice, most real-word graphs exhibit strong regularities
which make them far from the random and pathological cases.
Based on this remark, many geometry processing algorithms try to exploit this regularity
in order to obtain better results in practice.
For instance, when applied to regular graphs, many mesh compression schemes~\cite{Gotsman03} achieve 
good compression rates, well below the worst-case optimal bound guaranteed by~\cite{Pou03}.
As far as we know, the problem of providing an adaptive analysis of Schnyder woods taking into
account the graph regularity has not been investigated so far.
This work provides empirical evidence about the fact that balanced Schnyder woods can lead to fast solutions
achieving good results in practice, especially for real-world graphs.
As test applications, we evaluate the layout quality of a Schnyder drawing depending on the balance of the underlying Schnyder wood, and we consider
the problem of computing small separators for planar graphs, which has been extensively 
investigated~\cite{LiptonT1979,LiptonT80,Miller86,SpielmanT96}, due to its relevance for many graph algorithms.

\paragraph{Preliminaries and related works.}


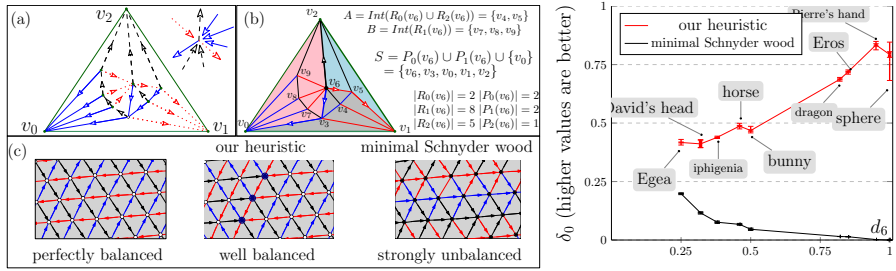
\begin{figure*}[t!]

\begin{minipage}{0.98\textwidth} 
\scalebox{0.44} {
\includegraphics{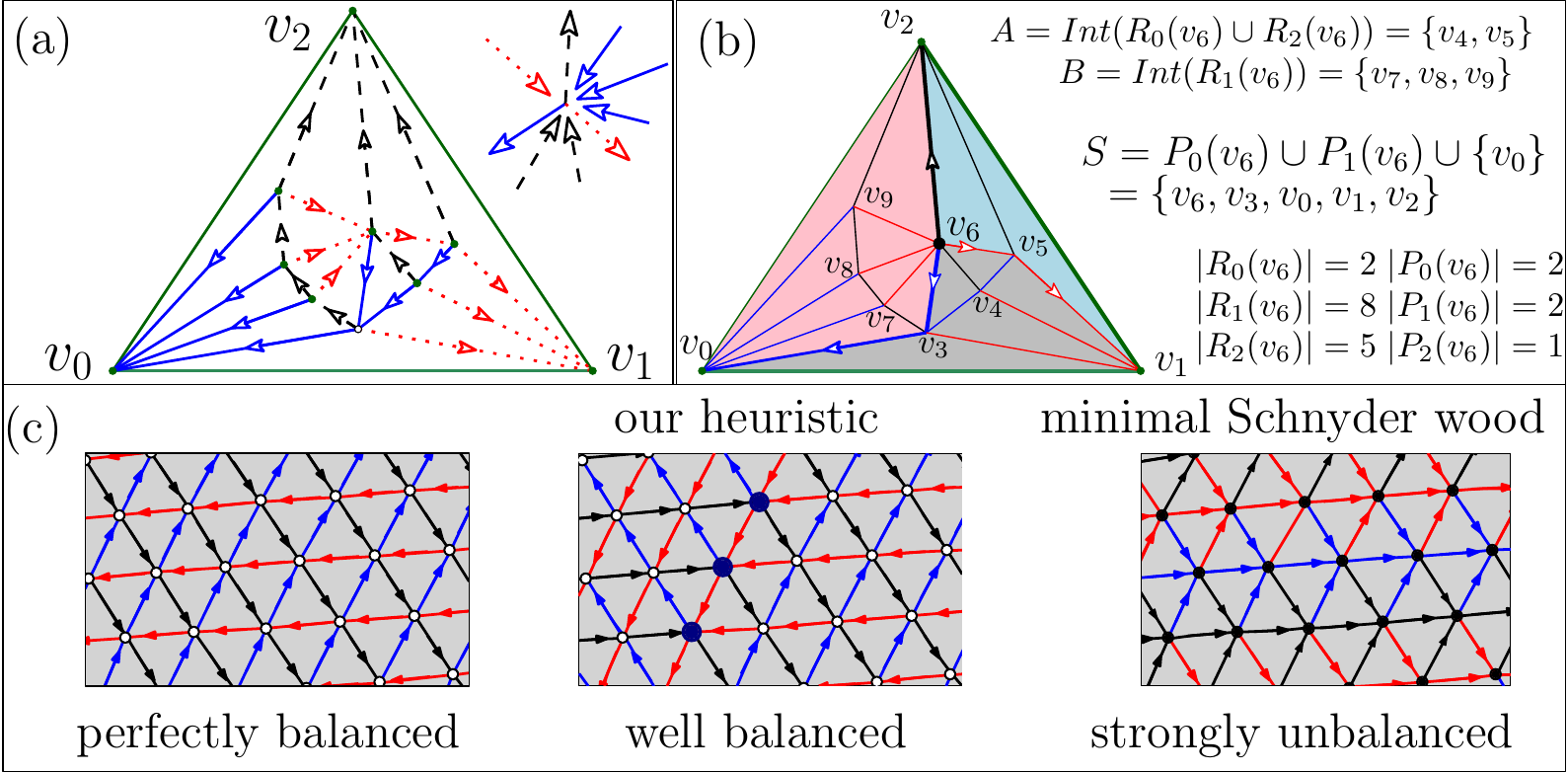}
}
\hspace{.46cm}
\scalebox{0.74}{
\begin{tikzpicture}

\pgfplotsset{width=8.8cm,compat=1.3}

\tikzset{
every pin/.style={fill=lightgray!50!white,rectangle,rounded corners=3pt,font=\small},
small dot/.style={fill=black,circle,scale=0.2}
}

\hspace{-1.25cm}
\scalebox{0.70} {
\begin{axis}[
    no markers,
    xlabel={\Large{$d_6$}},
    axis x line=middle,
    ylabel={\Large{\texttt{$\delta_0$} (higher values are better)}},
    separate axis lines=false,
    line width=0.5pt,
    xmin=0.0, xmax=1.01,
    ymin=0.0, ymax=1.0,
    xtick={0.25,0.5,0.75,1.0},
    ytick={0.,0.25,0.5,0.75,1},
    legend pos=north west,
    ymajorgrids=true,
    grid style=dashed,
]

\addplot [scatter, color=red, mark=none]
 plot [error bars/.cd, y dir = both, y explicit, error bar style={line width=2.5pt}]
 table[row sep=crcr, x index=0, y index=1, x error index=2, y error index=3,]{
0.25 0.417 0 0.002\\
0.32 0.411 0 0.016\\
0.38 0.439 0 0.001\\
0.46 0.487 0 0.003\\
0.5 0.468 0 0.003\\
0.82 0.687 0 0.002\\
0.85 0.719 0 0.002\\
0.95 0.834 0 0.005\\
0.999 0.793 0 0.012\\
};

\addplot [scatter, color=black, mark=none]
 plot [error bars/.cd, y dir = both, y explicit, error bar style={line width=2.5pt}]
 table[row sep=crcr, x index=0, y index=1, x error index=2, y error index=3,]{
0.25 0.198 0 0.001\\
0.32 0.116 0 0.001\\
0.38 0.076 0 0.0\\
0.46 0.067 0 0.0\\
0.5 0.046 0 0.0\\
0.82 0.015 0 0.0\\
0.85 0.014 0 0.0\\
0.95 0.002 0 0.0\\
0.999 0. 0. 0.\\
};

\addplot [scatter, color=red, only marks, mark=none]
 plot [error bars/.cd, y dir = both, y explicit, error bar style={line width=1pt}]
 table[row sep=crcr, x index=0, y index=1, x error index=2, y error index=3,]{
0.25 0.418 0 0.012\\ 
0.32 0.412 0 0.017\\ 
0.38 0.439 0 0.005\\ 
0.46 0.487 0 0.012\\ 
0.5 0.472 0 0.014\\ 
0.82 0.687 0 0.008\\ 
0.85 0.718 0 0.009\\ 
0.95 0.831 0 0.018\\ 
0.999 0.764 0 0.082\\ 
};
 
\addplot [scatter, color=black, only marks, mark=none]
 plot [error bars/.cd, y dir = both, y explicit, error bar style={line width=1pt}]
 table[row sep=crcr, x index=0, y index=1, x error index=2, y error index=3,]{
0.25 0.198 0 0.004\\
0.32 0.116 0 0.005\\
0.38 0.076 0 0.005\\
0.46 0.067 0 0.004\\
0.5 0.046 0 0.005\\
0.82 0.015 0 0.0\\
0.85 0.015 0 0.0\\
0.95 0.002 0 0.0\\
0.999 0. 0. 0.\\
};

\node[small dot,pin=260:{\large{Egea}}] at (axis description cs:0.24,0.38) {};
\node[small dot,pin=100:{\large{David's head}}] at (axis description cs:0.32,0.45) {};
\node[small dot,pin=270:{iphigenia}] at (axis description cs:0.38,0.42) {};
\node[small dot,pin=90:{\large{horse}}] at (axis description cs:0.46,0.52) {};
\node[small dot,pin=310:{\large{bunny}}] at (axis description cs:0.5,0.44) {};
\node[small dot,pin=265:{dragon}] at (axis description cs:0.81,0.66) {};
\node[small dot,pin=95:{\large{Eros}}] at (axis description cs:0.85,0.73) {};
\node[small dot,pin=110:{Pierre's hand}] at (axis description cs:0.94,0.86) {};
\node[small dot,pin=265:{\large{sphere}}] at (axis description cs:0.98,0.64) {};

\legend{\large{our heuristic}, \normalsize{minimal Schnyder wood}}

\end{axis}
}

\end{tikzpicture}
}
\end{minipage}

\caption{
\label{fig:defect}
(a) A planar triangulation endowed with a Schnyder wood. (b) a separator $(A, B, S)$ obtained from the Schnyder wood.
(c) three Schnyder woods of the same portion of a spherical grid: our heuristic 
leads to a majority of balanced vertices (white circles),
while the minimal Schnyder wood is strongly unbalanced.
(Right chart) Evaluation of the balance of Schnyder woods (tests are repeated with $500$ random seeds).
}
\end{figure*}


In this work we deal with \emph{planar triangulations}, which are genus $0$ simple maps where every face is triangulated (we will
denote by $n$ the number of vertices and by $m$ the number of edges).
Given a planar triangulation with a distinguished root (outer) face $(v_0, v_1, v_2)$, a Schnyder wood~\cite{Schnyder90} is defined
as a coloring (with colors $0$, $1$ or $2$) and orientation of the inner edges such that each inner vertex has exactly 
one outgoing incident edge for each color, and the remaining incident edges must satisfy the local Schnyder rule
(see Fig.~\ref{fig:defect}(a)).
A given rooted triangulation may admit many Schnyder woods~\cite{Bonichon09,Bonichon05,FelsnerZ08}: among them,
the \emph{minimal} one (without ccw oriented triangles) plays a fundamental role~\cite{Pou03,AleardiD18}.
Here we focus on balanced Schnyder woods, for which the ingoing edges are evenly distributed around inner vertices.
A related problem concerns the computation of egalitarian orientations: unfortunately the results in~\cite{BorradaileIMOWZ17}
only apply to unconstrained orientations.
%
Schnyder woods have led to a linear-time algorithm providing an elegant solution to the grid drawing problem
(solved independently also in~\cite{FPP90}): in its pioneristic work~\cite{Schnyder90}
Schnyder showed that a planar graph with $n$ vertices admits a straight-line drawing on a grid of size $O(n)\times O(n)$.
Schnyder drawings have a number of nice properties that make them useful for addressing problems~\cite{BonichonGHP10,BonichonGHI10,Dhandapani10} 
involving planar graphs in several distinct domains.
While recent works~\cite{LiSW17} provide a probabilistic study of the converge for uniformly sampled triangulations endowed with a Schnyder wood,
as far as we know there are no theoretical or empirical evaluations of the quality of Schnyder drawings for regular graphs.
%
Given a graph $G$ we consider \emph{small separators} which are defined by a partition $(A, B, S)$ of all vertices 
such that $S$ is a separating vertex set of small size (usually $|S|=O(\sqrt{m})$), and the remaining vertices in $G\setminus S$ belong
to a balanced partition $(A, B)$ satisfying $|A|\leq \alpha n$, $|B|\leq \alpha n$ (usually, for planar graphs, the \emph{balance ratio}
is $\alpha=\frac{2}{3}$).
Here we focus on \emph{simple cycle separators}~\cite{Miller86}, for which fast implementations~\cite{HolzerSWPZ09,FoxEpsteinMP016} 
have been recently proposed (some of them~\cite{FoxEpsteinMP016} are provided with a worst-case bound of $\sqrt{8m}$ on the cycle size).

\section{Contribution}

\subsection{Balanced Schnyder woods\label{sec:balancedSchnyderWoods}}

Our first step is to measure the balance of a Schnyder wood: given an inner vertex $v$ of degree $deg(v)$
having $indeg_{i}(v)$ incoming edges of color $i$ (for $i\in \{0, 1, 2\}$), 
we define its \emph{defect} as $\delta(v)=\max_{i} indeg_i(v) - \min_{i} indeg_{i}(v)$ if $deg(v)$ is a multiple of $3$, and
$\delta(v)=\max_{i} indeg_i(v) - \min_{i} indeg_{i}(v) - 1$ otherwise.
We say that a vertex is \emph{balanced} if $\delta(v)=0$
%
%
and a Schnyder wood is \emph{well balanced} if a majority of vertices have a small defect.
For regular graphs is possible, in principle, to get a Schnyder wood that is perfectly balanced ($\delta(v)=0$ everywhere)
as shown in Fig.~\ref{fig:defect}(c).
In practice many Schnyder woods are unbalanced 
and we are not aware of existing theoretical or empirical results on the balance of Schnyder woods.

\paragraph{An heuristic for well balanced Schnyder woods.}
%
%
We make use of the well known incremental vertex shelling procedure~\cite{Brehm_thesis} that
computes a Schnyder wood with a sequence of vertex removals.
This procedure has many degrees of freedom: at each step the choice of the vertex to be removed can possibly lead
to a different Schnyder wood.
%
%
In order to get as much as possible balanced vertices, we retard the removal of some vertices according
to a balance priority, defined as the total number of ingoing edges incident to a vertex during the shelling procedure.
The balance can be further improved by performing the reversal of oriented triangles in a post-processing~\footnote{
The results presented in Section~\ref{sec:evaluation} are obtained without post-processing step.
}
step.
\OnlyLongVersion{We refer to the Appendix for more details.}
\OnlyShortVersion{We refer to~\cite{...} for more details.}

\subsection{From Schnyder drawings to small simple cycle separators}

Schnyder woods provides a very fast procedure for partitioning, given an arbitrary inner vertex $v$, 
the set of inner faces of a triangulation into three sets $R_0(v)$, $R_1(v)$ and $R_2(v)$ (respectively blue, red and gray triangles in Fig.~\ref{fig:defect}(b)), 
whose boundaries consist of the three disjoint paths $P_0(v)$, $P_1(v)$ and $P_2(v)$ emanating from $v$.
The computation of simple cycle separators can be done as follows:
for each vertex $v$ check whether the two sets $A=Int(R_{i}(v)\cup R_{i+1}(v))$ and $B=Int(R_{i+2}(v))$
satisfy the prescribed balance ratio for at least one index $i\in\{0, 1, 2\}$ (indices are modulo $3$, and $Int(R)$ denotes the set of inner vertices of a region $R$):
then select the vertex for which the corresponding cycle length $|P_{i}(v)|+|P_{i+1}(v)|+1$ is minimal.
All this steps can be performed almost instantaneously, since all the quantities
above are encoded in the Schnyder drawing itself (see~\cite{Schnyder90} for more details).
As far as we know there are no theoretical guarantees on both the partition balance and boundary size: as observed in practice, 
most vertices lead to unbalanced partitions whose boundary size can be very large.
%

%


\begin{figure*}[t]
\hspace{0.8cm}
\scalebox{0.44}{

\begin{tikzpicture}

\tikzset{
every pin/.style={fill=lightgray!50!white,rectangle,rounded corners=3pt,font=\small},
small dot/.style={fill=black,circle,scale=0.25}
}

\begin{axis}[
    stack plots=y,
    area style,
    title={\large{Average timings (real-world graphs)}},
    xlabel={\large{$n$}},
    ylabel={\Large{seconds}},
    axis x line=middle,
    xmin=0.0, xmax=1.105,
    ymin=0.0, ymax=1.51,
    xtick={0.100000,0.500000,1.000000,1.100000,1.700000},
    ytick={0.5,1.0,1.5,2.0,2.5},
    legend pos=north west,
    ymajorgrids=true,
]

\addplot [fill=blue!40!white,mark=otimes]
coordinates {
(0.049922,0.0142) 
(0.174066,0.0465) 
(0.309506,0.0652) 
(0.476596,0.1004) 
(0.655980,0.144) 
(0.701322,0.1751) 
(0.773465,0.1531) 
(1.104470,0.2569) 
}
\closedcycle;

\addplot [fill=red!40!white]
coordinates {
(0.049922,0.0278)
(0.174066,0.0802)
(0.309506,0.1153)
(0.476596,0.1425) 
(0.655980,0.2054)
(0.701322,0.237)
(0.773465,0.2273)
(1.104470,0.3201) 
}
\closedcycle;

\addplot [fill=green!40!white,mark=otimes] 
    coordinates{
(0.049922,0.001)
(0.174066,0.005)
(0.309506,0.005)
(0.476596,0.005)
(0.655980,0.007)
(0.701322,0.008)
(0.773465,0.01)
(1.104470,0.013)
}
\closedcycle;

 \node[small dot,pin=90:{\large{Arc triomphe}}] at (axis description cs:0.16,0.14) {};
 \node[small dot,pin=90:{\large{Eros}}] at (axis description cs:0.42,0.19) {};
 \node[small dot,pin=100:{\normalsize{dragon}}] at (axis description cs:0.58,0.28) {};
 \node[small dot,pin=90:{\normalsize{Pierre's hand}}] at (axis description cs:0.7,0.34) {};
 \node[small dot,pin=120:{\large{Isidore}}] at (axis description cs:0.98,0.48) {};

 \legend{\large{step 1: balanced orientation},\large{step 2: Schnyder drawing},\large{step 3: shortest separator}}
 
\end{axis}


\hspace{7.6cm}
\begin{axis}[
    stack plots=y,
    area style,
    title={\large{Average timings (random triangulations)}},
    xlabel={\large{$n$}},
    axis x line=middle,
    xmin=0.0, xmax=2.000001,
    ymin=0.0, ymax=1.51,
    xtick={0.100000,0.500000,1.000000,1.500000,2.000000},
    ytick={0.5,1.0,1.5,2.0,2.5},
    legend pos=south west,
    ymajorgrids=true,
]

\addplot [fill=blue!40!white,mark=otimes]
coordinates {
(0.100000,0.0266) 
(1.000000,0.3268) 
(2.000000,0.6988) 
}
\closedcycle;

\addplot [fill=red!40!white]
coordinates {
(0.100000,0.0306) 
(1.000000,0.3009) 
(2.000000,0.5956) 
}
\closedcycle;

\addplot [fill=green!40!white,mark=otimes,mark=otimes] 
    coordinates{
(0.100000,0.0012) 
(1.000000,0.012) 
(2.000000,0.023) 
}
\closedcycle;

 \node[small dot,pin=80:{\large{R100k}}] at (axis description cs:0.05,0.08) {};
 \node[small dot,pin=100:{\Large{Random 1M}}] at (axis description cs:0.48,0.46) {};
 \node[small dot,pin=180:{\Large{Random 2M}}] at (axis description cs:0.98,0.88) {};

\end{axis}


\hspace{7.6cm}
\begin{axis}[
    no markers,
    title={\Large{total timing cost (steps 1+2+3)}},
    xlabel={\large{$n$}},
    axis x line=middle,
    xmin=0.0, xmax=2.02,
    ymin=0.0, ymax=1.51,
    xtick={0.100000,0.25,0.500000,0.75,1.000000,1.500000,2.000000},
    ytick={0.5,1.0,1.5,2.0,2.5},
    legend pos=south west,
    ymajorgrids=true,
]

\addplot [scatter, color=red, mark=none]
 plot [error bars/.cd, y dir = both, y explicit, error bar style={line width=2.5pt}]
 table[row sep=crcr, x index=0, y index=1, x error index=2, y error index=3,]{
0.1 0.045 0 0.001\\ 
0.25 0.128 0 0.002\\ 
0.5 0.261 0 0.005\\ 
0.75 0.427 0 0.01\\ 
1.0 0.565 0 0.013\\ 
1.5 0.952 0 0.017\\ 
2.0 1.255 0 0.024\\ 
};

\addplot [scatter, color=red, only marks, mark=none]
 plot [error bars/.cd, y dir = both, y explicit, error bar style={line width=1pt}]
 table[row sep=crcr, x index=0, y index=1, x error index=2, y error index=3,]{
0.1 0.059 0 0.018\\ 
0.25 0.154 0 0.034\\ 
0.5 0.269 0 0.02\\ 
0.75 0.462 0 0.062\\ 
1.0 0.577 0 0.051\\ 
1.5 0.97 0 0.075\\ 
2.0 1.27 0 0.103\\ 
};

 \node[small dot,pin=00:{\Large{globe500k}}] at (axis description cs:0.26,0.14) {};
 \node[small dot,pin=100:{\Large{globe750k}}] at (axis description cs:0.37,0.38) {};
 \node[small dot,pin=90:{\Large{globe1M}}] at (axis description cs:0.50,0.48) {};
 \node[small dot,pin=180:{\Large{globe2M}}] at (axis description cs:0.98,0.94) {};

\end{axis}

\end{tikzpicture}
}
\caption{
\label{fig:runtime}
Evaluation of timing costs over $100$ executions (allocating 1GB of RAM for the JVM): timings are expressed as a function of the size
(millions of vertices).
}
\end{figure*}
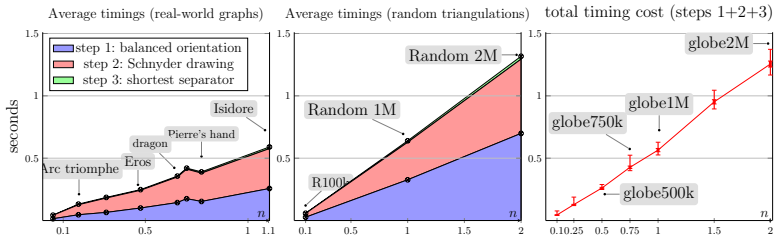

\subsection{Experimental results}\label{sec:evaluation}

\begin{figure*}[t!]

\begin{minipage}{1.0\textwidth} 
\scalebox{0.56}{
\input{Stats_separatorQuality}
}
\end{minipage}

\begin{center}
\scalebox{0.55} {
\includegraphics{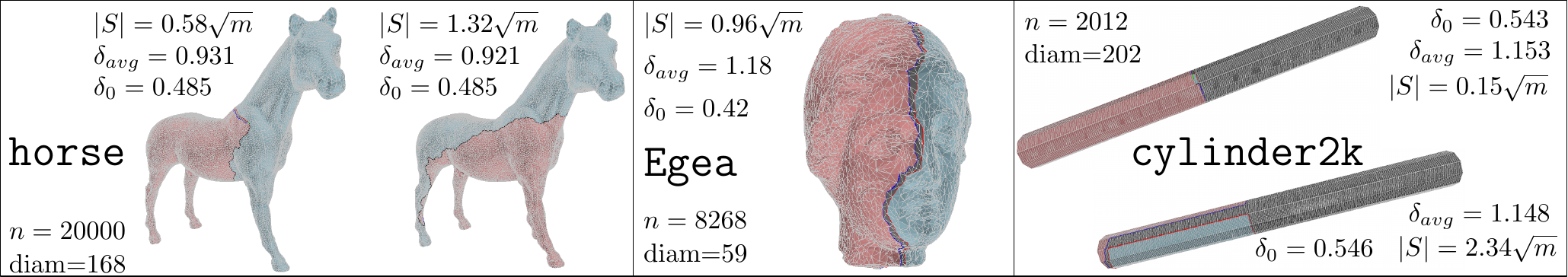}
}
\end{center}

\caption{
\label{fig:cycleLength}
We evaluate the quality of our simple cycle separators obtained from our balanced Schnyder woods (tests are repeated using $200$ random seeds). 
The left charts report the boundary sizes, 
while the right charts show the plots of the separator balance (the normalized size of the smallest of the two sets $A$ and $B$).
The graphs are listed from left to right according to the increasing values of their relative diameter.
}
\end{figure*}

\paragraph{Datasets and experimental setting.}
We run our experimental evaluations~\footnote{
Our datasets and code can be found at \tiny{\url{http://www.lix.polytechnique.fr/~amturing/software.html}}
} on a broad variety of graphs~\footnote{ 
Previous works~\cite{HolzerSWPZ09,FoxEpsteinMP016} triangulate the input graph
in a preprocessing phase.
},
including real-world meshes used in geometry processing 
(made available by the \texttt{aim@shape} and \texttt{Thingi10k} repositories), 
synthetic regular graphs with different shapes (\texttt{sphere}, \texttt{cylinder}, ...),
random planar triangulations (generated with uniform random sampling~\cite{Pou03}), and Delaunay triangulations
of random points.
As done in geometric modeling, we use the proportion of degree $6$ vertices, denoted by $d_6$, to measure
the regularity of a graph: $d_6$ is close to $1$ for regular meshes, while is usually below $0.3$ for irregular and random graphs.
To evaluate the balance of a Schnyder woods we use the proportion of balanced vertices, denoted by $\delta_{0}$, 
and the average defect computed on all vertices, denoted by $\delta_{avg}$.
As for previous works~\cite{HolzerSWPZ09,FoxEpsteinMP016}, the results (e.g. the size of the separator) 
can depend on the choice of the initial seed (the root face in our case).
We perform tests with hundreds of random seeds: for each choice of the seed,
we adopt whisker plots to show the entire range of computed values,
while each box represents the middle $50\%$ of values (as in Fig.~\ref{fig:cycleLength} and~\ref{fig:defect}).

\paragraph{Balance of Schnyder woods.}

To evaluate the balance quality of the Schnyder woods
we plot the value $\delta_0$ as a function of $d_6$: our balanced Schnyder woods are compared to minimal ones in Fig.~\ref{fig:defect}.
%
%
Experimental results strongly suggests that our heuristic leads to well balanced
Schnyder woods.
Our heuristic performs particularly well for regular graphs, for which a large majority of vertices are balanced ($79\%$ in average for the
\texttt{sphere} graph). The results are good also for irregular graphs (\texttt{egea}), where about $45\%$ of vertices are balanced.
Also observe that the choice of the initial seed has a limited effect on the balance of the resulting Schnyder wood.
Minimal Schnyder woods represent a bad case, especially for regular graphs:
most vertices have a large defect and the resulting paths $P_0(v)$, $P_1(v)$ and $P_2(v)$ resemble very long spirals.

\paragraph{Runtime performances.}

The algorithmic solutions relying on Schnyder woods are simple to implement
and extremely fast.
%
As observed in practice (see Fig.~\ref{fig:runtime}), our \texttt{Java} implementation allows processing
 between $1.43M$ and $1.92M$ vertices per second: we run our tests on an EliteBook with a core i7-5600U 2.60GHz 
 (with Ubuntu 16.04 and 1GB of RAM allocated for the JVM).
 This has to be compared to the \texttt{C} implementations of previous results on cycle separators~\cite{HolzerSWPZ09,FoxEpsteinMP016},
 running on an Intel Xeon X5650 2.67GHz (with $48.4$ GB of RAM): the fastest variant of the procedures tested in~\cite{FoxEpsteinMP016}
 allows processing between $0.54M$ and $0.62M$ vertices per second for the case of square \texttt{grids}.
 %
 %
 Our timing costs 
 are little affected by the choice of the initial seed and the structural properties of the graph.
 %
 Observe that once the Schnyder drawing is given,
 the extraction of the cycle separator is instantaneous ($0.01$ seconds for a $1M$ vertices graph).

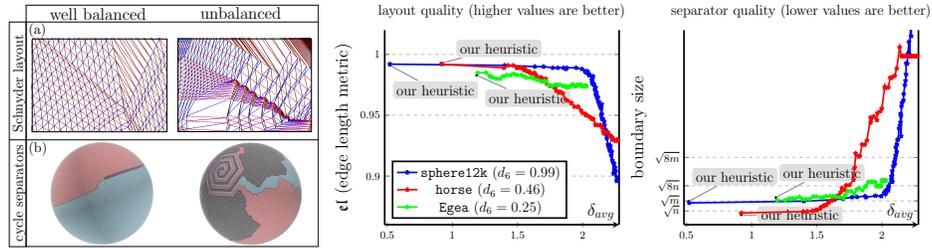
\begin{figure*}[t!]

\begin{minipage}{1.0\textwidth}
\hspace{-0.2cm}
\scalebox{0.98}{

\scalebox{0.54} {
\includegraphics{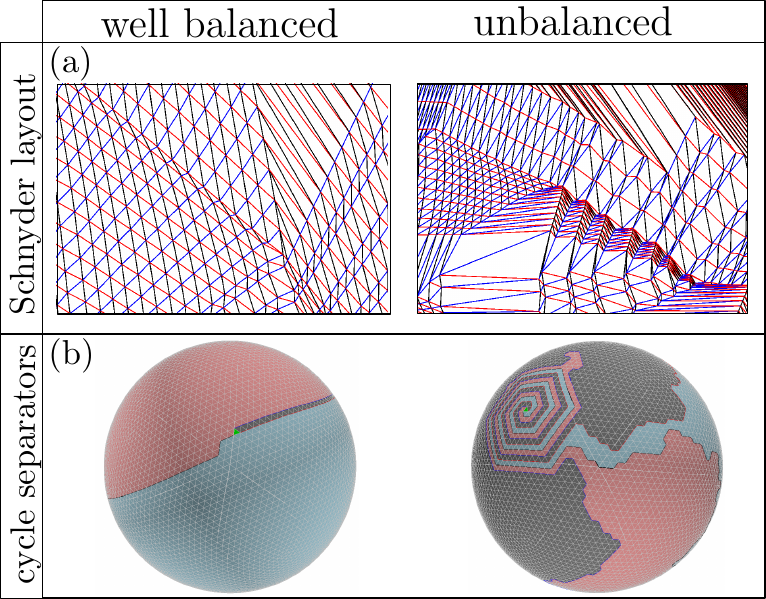}
}

\hspace{-0.4cm}
\scalebox{0.58}{
\begin{tikzpicture}

\tikzset{
every pin/.style={fill=lightgray!50!white,rectangle,rounded corners=3pt,font=\small},
small dot/.style={fill=black,circle,scale=0.25}
}

\scalebox{0.8} {
\begin{axis}[
    title={\large{layout quality (higher values are better)}},
    xlabel={\Large{$\delta_{avg}$}},
    axis x line=middle,
    ylabel={\Large{$\mathfrak{el}$ (edge length metric)}},
    line width=1.4pt,
    xmin=0.48, xmax=2.28,
    ymin=0.86, ymax=1.02,
    xtick={0.5,1.0,1.5,2.0},
    ytick={0.9,0.95,1.0},
    legend pos=south west,
    ymajorgrids=true,
    grid style=dashed,
]
 
 \node[small dot,pin=280:{\large{our heuristic}}] at (axis description cs:0.02,0.82) {};
 \node[small dot,pin=20:{\large{our heuristic}}] at (axis description cs:0.24,0.83) {};
 \node[small dot,pin=300:{\large{our heuristic}}] at (axis description cs:0.392,0.768) {};
 
\addplot[
    color=blue,
    mark=star,
    ]
    coordinates {
(0.52,0.992)
(1.4,0.991)(1.633,0.99)(1.7,0.99)(1.771,0.989)(1.81,0.989)(1.852,0.989)(1.871,0.988)
(1.878,0.989)(1.95,0.989)(1.994,0.988)(2.017,0.986)(2.029,0.985)(2.03,0.986)(2.038,0.986)(2.048,0.984)(2.05,0.983)(2.052,0.983)(2.075,0.98)(2.071,0.979)(2.073,0.977)(2.077,0.976)(2.079,0.976)
(2.08,0.982)(2.087,0.975)(2.094,0.971)(2.099,0.973)(2.101,0.967)(2.105,0.969)(2.106,0.967)(2.114,0.964)(2.115,0.965)(2.121,0.961)(2.124,0.961)(2.131,0.958)(2.138,0.956)(2.142,0.956)(2.146,0.951)
(2.143,0.95)(2.147,0.959)(2.15,0.947)(2.154,0.947)(2.165,0.945)(2.167,0.942)(2.176,0.938)(2.181,0.935)(2.185,0.933)(2.186,0.931)(2.194,0.926)(2.199,0.924)(2.207,0.921)(2.212,0.925)(2.219,0.917)(2.224,0.92)
(2.228,0.913)(2.236,0.911)(2.237,0.902)(2.241,0.91)(2.25,0.907)(2.253,0.904)(2.261,0.896)(2.262,0.899)(2.284,0.895)(2.3,0.894)(2.301,0.887)(2.318,0.886)
};

\addplot[color=red, mark=star]
    coordinates {
(0.919, 0.992)
(1.396,0.989)(1.427,0.991)(1.467,0.991)(1.494,0.989)(1.501,0.989)(1.524,0.989)(1.532,0.989)(1.55,0.987)(1.557,0.988)(1.575,0.986)(1.589,0.986)(1.594,0.985)(1.615,0.984)(1.629,0.985)(1.642,0.983)(1.646,0.981)
(1.67,0.981)(1.685,0.981)(1.692,0.978)(1.717,0.98)(1.725,0.976)(1.75,0.974)(1.782,0.973)(1.791,0.973)(1.798,0.97)(1.819,0.969)(1.831,0.969)(1.848,0.968)(1.868,0.963)(1.885,0.962)(1.913,0.96)(1.914,0.96)
(1.937,0.957)(1.962,0.957)(1.966,0.957)(1.988,0.954)(1.999,0.952)(2.022,0.951)(2.024,0.95)(2.049,0.949)(2.063,0.948)(2.071,0.947)(2.082,0.947)(2.098,0.942)(2.114,0.942)(2.136,0.942)(2.15,0.94)
(2.158,0.937)(2.162,0.935)(2.172,0.936)(2.188,0.936)(2.194,0.936)(2.205,0.934)(2.216,0.933)(2.225,0.932)(2.227,0.93)(2.241,0.93)(2.251,0.929)(2.264,0.93)(2.272,0.929)
};

\addplot[color=green, mark=star]
    coordinates {
(1.194,0.985)
(1.259,0.985)(1.287,0.982)(1.295,0.982)(1.313,0.981)(1.338,0.98)(1.367,0.98)(1.39,0.982)(1.416,0.983)(1.446,0.983)(1.471,0.984)(1.494,0.983)(1.528,0.983)(1.553,0.982)(1.57,0.981)(1.601,0.981)(1.625,0.98)(1.651,0.979)(1.68,0.978)(1.704,0.978)(1.719,0.978)(1.738,0.975)(1.763,0.974)(1.789,0.975)(1.805,0.975)(1.826,0.975)(1.851,0.974)(1.87,0.974)(1.876,0.973)(1.894,0.973)(1.906,0.974)(1.918,0.975)(1.931,0.975)(1.94,0.974)(1.95,0.974)(1.958,0.975)(1.969,0.975)(1.977,0.975)(1.983,0.975)(1.99,0.975)(1.993,0.974)(1.999,0.974)(2.001,0.974)(2.005,0.974)(2.01,0.974)(2.012,0.974)(2.013,0.974)(2.017,0.974)(2.018,0.974)(2.02,0.974)(2.022,0.974)(2.022,0.974)(2.023,0.974)(2.024,0.974)(2.024,0.974)(2.024,0.974)(2.025,0.974)(2.026,0.974)(2.026,0.974)(2.027,0.974)(2.027,0.974)
};

\legend{\large{\texttt{sphere12k} ($d_6=0.99$)},\large{\texttt{horse} ($d_6=0.46$)},\large{\texttt{Egea} ($d_6=0.25$)}}
 
\end{axis}
}


\hspace{7cm}
\scalebox{0.80} {
\begin{axis}[
    title={\large{separator quality (lower values are better)}},
    xlabel={\Large{$\delta_{avg}$}},
    axis x line=middle,
    ylabel={\Large{boundary size}},
    line width=1.4pt,
    xmin=0.48, xmax=2.28,
    ymin=0.0, ymax=5.0,
    xtick={0.5,1,1.5,2.,2.5},
    ytick={0.353,0.612,1.,1.73}, 
    yticklabels={
    $\sqrt{n}$, 
    $\sqrt{m}$, 
    $\sqrt{8n}$, 
    $\sqrt{8m}$}, 
    legend pos=north west,
    ymajorgrids=true,
    grid style=dashed,
]
 
 \node[small dot,pin=80:{\large{our heuristic}}] at (axis description cs:0.02,0.12) {};
 \node[small dot,pin=0:{\large{our heuristic}}] at (axis description cs:0.244,0.05) {};
 \node[small dot,pin=80:{\large{our heuristic}}] at (axis description cs:0.392,0.14) {};

\addplot[color=blue,mark=star]
    coordinates {
(0.52,0.56)
(1.4,0.605)(1.633,0.642)(1.7,0.667)(1.771,0.708)(1.81,0.714)(1.871,0.735)
(1.878,0.704)(1.95,0.745)(1.994,0.732)(2.017,0.822)(2.029,0.875)(2.03,0.798)(2.038,0.897)(2.048,0.878)(2.05,0.993)(2.052,0.965)(2.071,1.158)(2.073,1.239)(2.075,1.102)(2.077,1.257)(2.079,1.381)(2.08,1.105)
(2.087,1.506)(2.094,1.611)(2.099,1.565)(2.101,1.689)(2.105,1.785)(2.106,1.993)(2.114,1.869)(2.115,2.061)(2.121,2.183)(2.124,2.198)(2.131,2.475)(2.138,2.546)(2.142,2.633)(2.143,2.853)(2.146,2.797)
(2.147,2.276)(2.15,3.043)(2.154,3.133)(2.165,3.195)(2.167,3.36)(2.176,3.589)(2.181,3.701)(2.185,3.863)(2.186,4.145)(2.194,4.142)(2.199,4.397)(2.207,4.698)(2.212,4.375)(2.219,5.049)(2.224,4.844)
(2.228,5.176)(2.236,5.422)(2.237,6.064)(2.241,5.477)(2.25,5.633)(2.253,5.959)(2.261,6.446)(2.262,6.173)(2.284,6.583)(2.3,6.602)(2.301,6.949)(2.318,7.164)
};

\addplot[color=red,mark=star]
    coordinates {
(0.919,0.302)
(1.396,0.337)(1.427,0.34)(1.467,0.355)(1.494,0.38)(1.501,0.37)(1.524,0.407)(1.532,0.452)(1.55,0.477)(1.557,0.497)(1.575,0.512)(1.589,0.567)(1.594,0.577)(1.615,0.62)(1.629,0.647)(1.642,0.707)(1.646,0.725)
(1.67,0.767)(1.685,0.845)(1.692,0.947)(1.717,0.887)(1.725,1.012)(1.75,1.017)(1.782,1.152)(1.791,1.467)(1.798,1.835)(1.819,1.75)(1.831,1.765)(1.848,2.06)(1.868,2.055)(1.885,2.165)(1.913,2.757)(1.914,2.63)
(1.937,2.785)(1.962,2.687)(1.966,2.87)(1.988,2.915)(1.999,2.99)(2.022,3.547)(2.024,3.512)(2.049,3.527)(2.063,3.777)(2.071,3.77)(2.082,3.802)(2.098,3.827)(2.114,4.017)(2.136,4.562)(2.15,4.332)
(2.158,4.332)(2.162,4.332)(2.172,4.332)(2.188,4.332)(2.194,4.332)(2.205,4.332)(2.216,4.332)(2.225,4.332)(2.227,4.332)(2.241,4.332)(2.251,4.332)(2.264,4.332)(2.272,4.332)
};

\addplot[color=green,mark=star]
    coordinates {
(1.194,0.626)
(1.259,0.629)(1.287,0.672)(1.295,0.657)(1.313,0.672)(1.338,0.676)(1.367,0.715)(1.39,0.657)(1.416,0.657)(1.446,0.657)(1.471,0.719)(1.494,0.723)(1.528,0.707)(1.553,0.699)(1.57,0.73)(1.601,0.738)(1.625,0.734)(1.651,0.715)(1.68,0.707)(1.704,0.684)(1.719,0.769)(1.738,0.867)(1.763,0.867)(1.789,0.793)(1.805,0.902)(1.826,0.937)(1.851,0.886)(1.87,0.979)(1.876,0.983)(1.894,1.096)(1.906,1.096)(1.918,1.1)(1.931,1.1)(1.94,1.096)(1.95,1.092)(1.958,0.987)(1.969,0.991)(1.977,0.995)(1.983,0.991)(1.99,0.995)(1.993,0.995)(1.999,1.139)(2.001,1.139)(2.005,1.139)(2.01,1.139)(2.012,1.139)(2.013,1.139)(2.017,1.139)(2.018,1.139)(2.02,1.139)(2.022,1.139)(2.022,1.139)(2.024,1.139)(2.024,1.139)(2.024,1.139)(2.023,1.139)(2.025,1.139)(2.026,1.139)(2.026,1.139)(2.027,1.139)(2.027,1.139)
};

 
\end{axis}
}

\end{tikzpicture}
}

}
\end{minipage}

\caption{
\label{fig:layout+separators}
For a fixed initial seed, we generate a sequence of Schnyder woods by starting from a well balanced Schnyder wood (computed with our heuristic)
and by randomly reversing ccw oriented triangles.
In the charts we plot the layout and separator quality as functions of the average defect $\delta_{avg}$ of the corresponding Schnyder wood. 
}
\end{figure*}


\paragraph{Layout quality.}
A qualitative evaluation of the graph layouts based on the balance Schnyder woods is provided by
the pictures in Fig.~\ref{fig:layout+separators}(a) showing two portions of the Schnyder drawings of a regular sphere graph.
When starting from a our well balanced Schnyder woods the shape of triangles is much
more balanced, and the resulting drawing partially captures the regularity of the grid.
When starting from an unbalanced Schnyder wood the drawing exhibits many long edges and flat triangles, a typical
drawback of Schnyder drawings.
In order to provide a quantitative measure of the layout quality we consider the \emph{edge lengths aesthetic metric}
defined by $\mathfrak{el}=1-d_{el}$, where $d_{el}$ is the average percent deviation of edge lengths: values
close to $1$ mean that most edges have the same length (see~\cite{FowlerK12} for more details).
For a fixed initial seed, we start from a balanced Schnyder wood obtained with our heuristic 
and we randomly reverse ccw oriented triangles,
obtaining a sequence of Schnyder woods which are more and more unbalanced.
%
The middle chart in Fig.~\ref{fig:layout+separators} reports the values of $\mathfrak{el}$ as a function of the average defect:
the layout quality tends to deteriorate as soon as Schnyder woods get more unbalanced (high values of $\delta_{avg}$).

\paragraph{Length and balance of separators.}


We look for separators with a balance ratio $\alpha=\frac{2}{3}$ that are \emph{short}: 
the boundary size is at most $|S|\leq \sqrt{8m}$, as required in~\cite{FoxEpsteinMP016}.
We plot in the charts of Fig.~\ref{fig:cycleLength} the boundary sizes and partition balances
of the separators obtained from a Schnyder drawing as described in Section~\ref{sec:balancedSchnyderWoods}.
Our tests, repeated over several tens of graphs, confirm our intuition that balanced Schnyder woods lead to good separators
for a large majority of classes of graphs. As for the layout quality, the separator size and balance strongly depend on the balance of the underlying
Schnyder wood (right chart in Fig.~\ref{fig:layout+separators}).
%
The boundary size of the separator is affected by the choice of the seed
for graphs with large diameter: a good choice of the seed would prevent from getting too long cycles.
For graphs with small diameter (e.g. random triangulations) Schnyder woods lead to very short separators,
while the size is closed to $\sqrt{m}$ for most real-world graphs. Our separators are often longer when compared
with the results obtained in~\cite{FoxEpsteinMP016}, but well below the prescribed bound of $\sqrt{8m}$.

%
%
%

\bibliographystyle{splncs04}
\bibliography{GraphDrawing}

\OnlyLongVersion{\appendix
\section{Appendix: computation of balanced Schnyder woods}\label{sec:appendix}

\subsection{Incremental vertex conquest}

\begin{figure}[th!]
  \includegraphics[width=\textwidth]{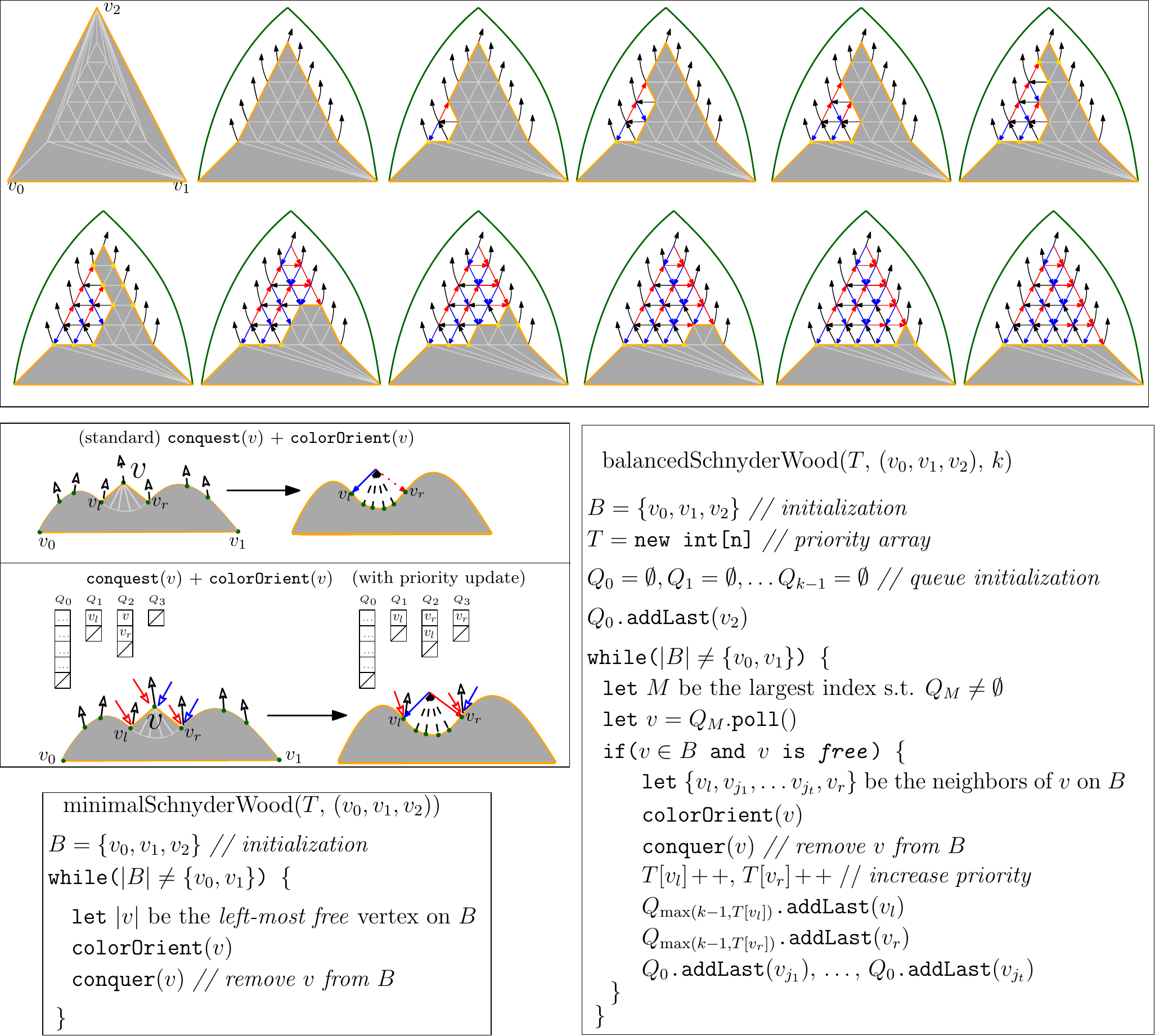}
 \caption{The pictures above illustrate some steps of our heuristic for computing a balanced Schnyder wood of a planar triangulation. Yellow circles represent boundary
 vertices incident to chordal edges (they are not free and cannot be removed).
\label{fig:computation}}
\end{figure}

For the sake of completeness we first recall the incremental procedure
that allows us to obtain a Schnyder wood of a planar triangulation $G$: we follow the approach 
based on vertex shellings as described in~\cite{Brehm_thesis,Kobourov16}.

The main idea is to incrementally maintain a region $C$ homeomorphic to a topological disk (white faces in Fig.~\ref{fig:computation}):
its boundary is given by a simple cycle $B:=\{v_0, v_{j_1}, v_{j_2}, \ldots , v_{j_k}, v_1\}$
(orange edges in Fig.~\ref{fig:computation}).
$C$ is initialized to be the root (outer) face, and its boundary $B$ consists of the outer vertices 
$(v_0, v_1, v_2)$.
The algorithm performs a sequence of $n-2$ conquest of vertices lying on $B$ until the boundary is reduced to $(v_0, v_1)$.
An important remark is that only \emph{free} vertices on $B$ can be removed: 
we avoid to remove vertices having incident chordal edges (drawn as light gray segments),
so that the cycle $B$ remains simple at each step (no self-intersections, the interior being simply connected).
By definition the \texttt{conquest} of a free vertex $v$ ($v$ is different from $v_0$ and $v_1$) 
consists in removing $v$ from $B$, together with all its incident faces.
Together with the removal of a vertex, we perform a \texttt{colorOrient} operation: 
given a vertex $v$ on $B$, with left neighbor $v_l$ and right neighbor $v_r$ on the cycle $B$, 
we assign color $0$ to the edge $(v_r, v)$ and
color $1$ to the edge $(v_l, v)$, both being oriented outgoing from vertex $v$.
All remaining possibly existing edges incident to $v$ are assigned color $2$ and oriented toward $v$.
Using an induction argument it is easy to see that the algorithm above always terminates, and the resulting edge orientation
satisfy the definition of Schnyder woods (we refer to~\cite{Brehm_thesis} for a more detailed presentation).

\paragraph{Minimal (maximal) Schnyder wood.} As one can observe, at each step there are many choices among the possible free vertices on $B$,
which possibly leads to many distinct Schnyder woods.
Among all Schnyder woods, the minimal (resp. maximal) one plays an important role in many algorithmic and combinatorial problems.
A fundamental property is that the minimal (resp. maximal) Schnyder wood has no ccw (resp. cw) oriented cycles of directed edges.
The computation of the minimal (resp. maximal) Schnyder wood can be performed as before:
at each step just perform the conquest of the free vertex $v$ on $B$ that is the closest
to $v_0$ (resp. to $v_1$).

\subsection{A retarded vertex conquest for computing balanced Schnyder woods}

In order to get as much as possible balanced vertices, we retard the removal of some vertices according
to a balance priority.
At a given step of the shelling procedure, we define the \emph{priority} of a vertex $v$ as the total number incoming incident edges
that already have been assigned a color and orientation (toward $v$).

Then we make use of very simple (truncated) priority rule: among the free vertices on $B$ we select one with maximal priority to be removed.
If there are several boundary free vertices with the same priority, we remove the oldest vertex (the one that was added first to the cycle $B$).
Intuitively, the goal of this approach is to retard the conquest of vertices having a small number of ingoing edges:
observe that removing a vertex $v$ having $0$ ingoing edges leads to get $d-3$ edges of color $2$ ingoing at $v$, while
the number of ingoing edges of colors $0$ and $1$ would remain $0$ (thus $v$ would be unbalanced when its degree is greater than $4$).
From the implementation point of view, we use a collection of $k$ queues $Q_0, Q_2, \ldots, Q_{k-1}$ (where $k$ is a small constant),
to store the vertices according to their priorities (the priority of vertices is stored and maintained using an integer array $P$ of size $n$).

At the beginning of execution $Q_0$ contains only $v_2$ and the remaining queues are empty.
When performing the vertex conquest of a free vertex $v$ on $B$ we add to $Q_0$ the neighboring vertices that are discovered (getting an outgoing black edge),
while we increase the priority of $v_l$ and $v_r$, the two left and right neighboring vertices of $v$, since they receive a new incoming edge (of color $0$ and $1$ respectively).
We then add $v_l$ and $v_r$ at the end of $Q_{P[v_l]}$ and $Q_{P[v_r]}$ respectively.
We do not increase the priority of a vertex having priority $k-1$ (thus having $k-1$ ingoing edges): the vertex is added to the end of $Q_{k-1}$.
Observe that each vertex is added exactly once to $Q_0$, and can appear at most in $Q_0, Q_1, Q_2, \ldots Q_{r}$, where $r=\min (k-1, deg(v)-3)$, since
we have $k$ queues, and a vertex receives $deg(v)$ incoming edges.
At a given iteration of the incremental conquest, in order to chose a vertex candidate to be removed, we look for the largest index $M$ for which $Q_M$ is not empty:
we remove the first vertex $u$ from the front of $Q_M$ and we check whether $u$ is still on $B$ and free (not incident chords), and we perform a $\texttt{colorOrient}$
operation if this is the case.

This procedure terminates since there is at least one free vertex $u$ on $B$ at each step of the incremental shelling (see~\cite{Brehm_thesis} for more details),
and $u$ must belong to at least one of the queues: when a vertex $u\in B$ becomes free (because of the removal of an incident chordal edge) then its priority increases and $u$
is added to a new queue with higher.
Our heuristic is very simple to implement and its linear-time complexity is due to the fact that each vertex is added and removed from the queues
at most $deg(v)-3$ times.

The linear-time behavior is confirmed by our experiments, where we set $k=5$ (see plots in Fig.~\ref{fig:runtime}): the computation of balanced Schnyder woods is slower than
the computation of minimal Schnyder woods, but it is still extremely fast in practice, allowing us to process more than $3M$ vertices per second in the case of random triangulations.
It is even faster on mesh graphs, because of the large number of low degree vertices.

\subsection{Post-processing: improving the balance}

\begin{figure*}[t!]

\begin{minipage}{1.0\textwidth} 
\scalebox{1.0}{
\begin{tikzpicture}

\pgfplotsset{width=8.8cm,compat=1.3}

\tikzset{
every pin/.style={fill=lightgray!50!white,rectangle,rounded corners=3pt,font=\small},
small dot/.style={fill=black,circle,scale=0.2}
}

\hspace{-0.5cm}
\scalebox{0.70} {
\begin{axis}[
    title={\large{Real-world meshes}},
    xlabel={\Large{$d_6$}},
    axis x line=middle,
    ylabel={\Large{\texttt{$\delta_0$} (higher values are better)}},
    separate axis lines=false,
    line width=0.5pt,
    xmin=0.0, xmax=1.01,
    ymin=0.0, ymax=1.64,
    xtick={0.25,0.5,0.75,1.0},
    ytick={0.,0.25,0.5,0.75,1},
    legend pos=north west,
    ymajorgrids=true,
    grid style=dashed,
]
 
 \node[small dot,pin=260:{\large{Egea}}] at (axis description cs:0.238,0.40) {};
 \node[small dot,pin=100:{\large{David's head}}] at (axis description cs:0.32,0.452) {};
 \node[small dot,pin=270:{iphigenia}] at (axis description cs:0.38,0.48) {};
 \node[small dot,pin=90:{\large{horse}}] at (axis description cs:0.458,0.52) {};
 \node[small dot,pin=310:{\large{bunny}}] at (axis description cs:0.5,0.50) {};
 \node[small dot,pin=100:{dragon}] at (axis description cs:0.808,0.56) {};
 \node[small dot,pin=70:{\large{Eros}}] at (axis description cs:0.84,0.56) {};
 \node[small dot,pin=270:{hand}] at (axis description cs:0.94,0.54) {};

\addplot [color=green,mark=diamond*]
coordinates {
(0.25,0.70)(0.32,0.74)(0.38,0.81)(0.46,0.84)(0.5,0.83)(0.82,0.88)(0.85,0.88)(0.95,0.93)
}; 

\addplot [color=red,mark=square*]
coordinates {
(0.25,0.42)(0.32,0.41)(0.38,0.439)(0.46,0.487)(0.5,0.47)(0.82,0.59)(0.85,0.70)(0.95,0.81)
}; 

\addplot [color=black,mark=otimes*]
coordinates {
(0.25,0.19)(0.32,0.11)(0.38,0.07)(0.46,0.06)(0.5,0.046)(0.82,0.015)(0.85,0.015)(0.95,0.002)
}; 

\legend{\large{our heuristic+post-processing}, \large{our heuristic}, \normalsize{minimal Schnyder wood}}

\end{axis}
}

\hspace{6cm}
\scalebox{0.70} {
\begin{axis}[
    title={\large{Synthetic and random graphs}},
    xlabel={\Large{$d_6$}},
    axis x line=middle,
    separate axis lines=false,
    line width=0.5pt,
    xmin=0.0, xmax=1.01,
    ymin=0.0, ymax=1.64,
    xtick={0.25,0.5,0.75,1.0},
    ytick={0.,0.25,0.5,0.75,1},
    legend pos=north west,
    ymajorgrids=true,
    grid style=dashed,
]
 
 \node[small dot,pin=90:{\large{random100k}}] at (axis description cs:0.112,0.348) {};
 \node[small dot,pin=350:{\large{Delaunay100k}}] at (axis description cs:0.29,0.42) {};
 \node[small dot,pin=110:{\large{sphere}}] at (axis description cs:0.984,0.56) {};

\addplot [color=green,mark=diamond*]
coordinates {
(0.11,0.57)(0.29,0.74)(0.99,0.91)
}; 

\addplot [color=red,mark=square*]
coordinates {
(0.11,0.51)(0.29,0.40)(0.99,0.82)
}; 

\addplot [color=black,mark=otimes*]
coordinates {
(0.11,0.47)(0.29,0.14)(0.999,0.001)
}; 

\legend{\large{our heuristic+post-processing}, \large{our heuristic}, \normalsize{minimal Schnyder wood}}

\end{axis}
}

\end{tikzpicture}
}
\end{minipage}

\caption{
\label{fig:postprocessingDefect}
Post-processing phase: these charts show the effect of the post-processing phase that consists in reversing oriented triangles in order
to increase the number of balanced vertices. For a given fixed choice of the initial seed we evaluate the number of balanced vertices
that we obtain applying the post-processing phase after running our heuristic; we show a comparison with the balance of minimal Schnyder woods.
}
\end{figure*}
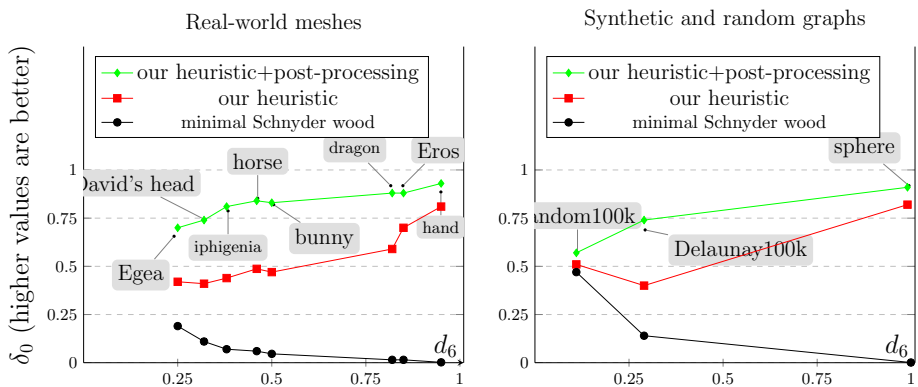


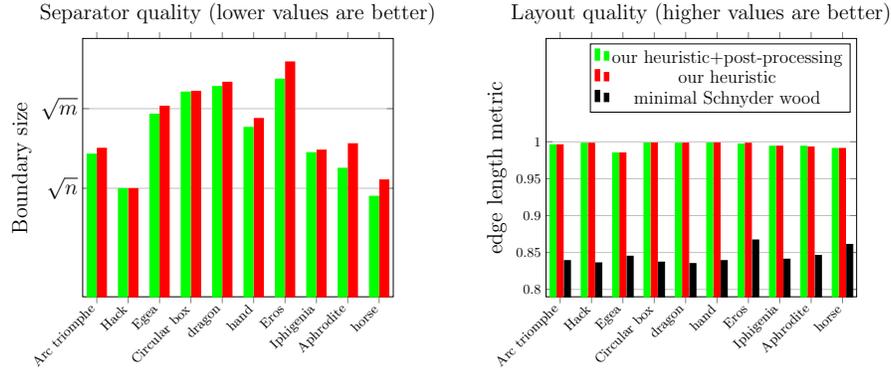
\begin{figure*}[t!]

\begin{minipage}{1.0\textwidth} 
\scalebox{0.88}{
\begin{tikzpicture}

\pgfplotsset{width=8.8cm,compat=1.3}

\tikzset{
every pin/.style={fill=lightgray!50!white,rectangle,rounded corners=3pt,font=\small},
small dot/.style={fill=black,circle,scale=0.2}
}

\hspace{-0.5cm}
\scalebox{0.65} {
\begin{axis}[
     title={\Large{Separator quality (lower values are better)}},
     ylabel={\Large{Boundary size}},
     xtick=\empty,
     ytick={0.353,0.612,1.,1.73}, 
     yticklabels={
     \Large{$\sqrt{n}$}, 
     \Large{$\sqrt{m}$}, 
     \Large{$\sqrt{8n}$}, 
     \Large{$\sqrt{8m}$}}, 
        ybar=2*\pgflinewidth,
        bar width=6pt,
        ymajorgrids = true,
        symbolic x coords={Arc triomphe,Hack,Egea,Circular box,dragon,hand,Eros,Iphigenia,Aphrodite,horse},
        xtick = data,
        x tick label style={rotate=45,anchor=east},
        scaled y ticks = false,
        enlarge x limits=0.05,
        ymin=0,
        legend cell align=left,
        legend style={
                at={(1,1.05)},
                anchor=south east,
                column sep=1ex
        }
]

 \addplot[style={green,fill=green,mark=none}]
  coordinates {
  (Arc triomphe,0.464)(Hack,0.352)(Egea,0.594)(Circular box,0.666)(dragon,0.684)(hand,0.551)(Eros,0.708)(Iphigenia,0.469)(Aphrodite,0.418)(horse,0.327)
  };
70
 \addplot[style={red,fill=red,mark=none}]
  coordinates {
  (Arc triomphe,0.483)(Hack,0.352)(Egea,0.62)(Circular box,0.668)(dragon,0.698)(hand,0.580)(Eros,0.764)(Iphigenia,0.477)(Aphrodite,0.498)(horse,0.38)
  };


\end{axis}
}

\hspace{7cm}
\scalebox{0.65} {
\begin{axis}[
     title={\Large{Layout quality (higher values are better)}},
     ylabel={\Large{edge length metric}},
     ymin=0.79, ymax=1.14,
     xtick=\empty,
     ytick={0.8,0.85,0.9,0.95,1.}, 
        ybar=2*\pgflinewidth,
        bar width=4pt,
        ymajorgrids = true,
        symbolic x coords={Arc triomphe,Hack,Egea,Circular box,dragon,hand,Eros,Iphigenia,Aphrodite,horse},
        xtick = data,
        x tick label style={rotate=45,anchor=east},
        scaled y ticks = false,
        enlarge x limits=0.05,
]

 \addplot[style={green,fill=green,mark=none}]
  coordinates {
  (Arc triomphe,0.996)(Hack,0.998)(Egea,0.985)(Circular box,0.9983)(dragon,0.9980)(hand,0.9984)(Eros,0.997)(Iphigenia,0.994)(Aphrodite,0.994)(horse,0.991)
  };

 \addplot[style={red,fill=red,mark=none}]
  coordinates {
  (Arc triomphe,0.996)(Hack,0.998)(Egea,0.985)(Circular box,0.9983)(dragon,0.9981)(hand,0.9985)(Eros,0.998)(Iphigenia,0.994)(Aphrodite,0.993)(horse,0.991)
  };

 \addplot[style={black,fill=black,mark=none}]
  coordinates {
  (Arc triomphe,0.839)(Hack,0.836)(Egea,0.845)(Circular box,0.837)(dragon,0.835)(hand,0.839)(Eros,0.867)(Iphigenia,0.841)(Aphrodite,0.846)(horse,0.861)
  };

\legend{\large{our heuristic+post-processing}, \large{our heuristic}, \large{minimal Schnyder wood}}

\end{axis}
}

\end{tikzpicture}
}
\end{minipage}

\caption{
\label{fig:postprocessingLayoutSeparator}
Effect of the post-processing phase on the quality of the Schnyder layout and on the size of the separators. All results are obtained
for a fixed choice o the initial seed.
}
\end{figure*}


The balance of the Schnyder woods obtained via our heuristic can be further improved by performing the reversal of oriented triangles.
To be more precise, one can iterate over all (cw or ccw) oriented triangle faces and check whether the reversal of the orientation
leads to increase the number of balanced vertices incident to the triangle: if this occurs the reversal is performed leading to
a Schnyder wood with slightly larger value of $\delta_0$.
As illustrated by the charts in Fig.~\ref{fig:postprocessingDefect}, the improvement one can achieve with this post-processing phase
is not negligible for regular graphs: as reported in the left chart the vertex balance has increased between
$14\%$ and $88\%$ on the tested meshes. The improvement is negligible for irregular (random) graphs (see right chart).

Unfortunately the computational cost of the post-processing phase is far from being negligible: according to our preliminary experiments,
iterating over the faces and reversing oriented triangles can be even $5.5$ slower than computing a balanced Schnyder wood with our heuristic.

Finally, observe that the layout and separator quality of the resulting Schnyder woods do not considerably improve after the post-processing phase
(see charts in Fig.~\ref{fig:postprocessingLayoutSeparator}).}

\end{document}